\documentclass[12pt]{article}
\usepackage[english]{babel}
\usepackage{graphicx}
\usepackage{amsmath}
\usepackage{amssymb,color}

\usepackage{bbm} 
\newcommand{\be}{\begin{equation}}
\newcommand{\ee}{\end{equation}}
\newcommand{\bqa}{\begin{eqnarray}}
\newcommand{\eea}{\end{eqnarray}}
\newcommand{\beq}{\begin{equation}}
\newcommand{\eeq}{\end{equation}}
\newcommand{\bea}{\begin{eqnarray}}

\newcommand{\non}{\nonumber}

\def\de{\partial}
\def\Tr{ \hbox{\rm Tr}}

\def\brc{\langle}
\def\ckt{\rangle}
\def\de{\partial}

\begin{document}

\thispagestyle{empty}
\begin{flushright}
IFUP-TH/2010-10\\
\end{flushright}
\vspace{3mm}

\begin{center}

{\Large \bf   New Results on Non-Abelian Vortices \\ 

\vspace{2mm}

 --  further insights into monopole,  vortex and confinement } \footnote{Talk presented at the 2009 International Workshop on
``Strong Coupling Gauge Theories in LHC Era'' [SCGT 09],    
 December 8-11, 2009, 
    Nagoya University,   Japan}
\\[15mm]

{K. KONISHI$^*$}

{Department of Physics, ``E. Fermi'',  University of Pisa, and   \\ INFN, Sezione  di Pisa\\
Largo Pontecorvo, 3, 56127, Pisa, Italy \\
$^*$e-mail: konishi@df.unipi.it\\
http://www.df.unipi.it/~konishi/   }



\vspace{6mm}

\vspace{6mm}

\vspace{6mm}

{\bf ABSTRACT }\\[5mm]

{\parbox{13cm}{\hspace{5mm}

We discuss some of the latest results concerning the non-Abelian vortices.  
The first concerns the construction of non-Abelian BPS vortices based on general gauge groups of the form $G= G^{\prime} \times U(1)$.  In particular detailed results about the vortex moduli space  have been obtained for  $G^{\prime}=SO(N)$ or $USp(2N)$.  The second result is about the ``fractional vortices'',  i.e., vortices of the minimum winding but having 
substructures in the tension (or flux) density in the transverse plane.  Thirdly, we discuss briefly the monopole-vortex complex.

}}
\end{center}

\newpage

\section{Introduction}

The last few years have witnessed a remarkable progress in our understanding of the non-Abelian vortices and their relation to monopoles, both of which are 
thirty-year old problems in theoretical physics, and which can bear important implications to some deep issues such as quark confinement. 
The plan of this talk is:  (i) a very brief review  of  non-Abelian monopoles; (ii) a brief review of  '03-'07 results on non-Abelian vortices; (iii) a new result on non-Abelian vortices based on general gauge groups; (iv) the fractional vortices; and (v) a brief discussion on the monopole-vortex complex and non-Abelian duality. 

It has become customary to think of quark confinement as a dual superconductor, in which (chromo-) electric charges are confined in a medium in which  a magnetic charge is condensed. The original suggestion by 't Hooft and Mandelstam is essentially Abelian: the effective low-energy degrees of freedom are 
magnetic monopoles arising from the Yang-Mills gauge fields.  It is however possible that the dual superconductor relevant to quark confinement is of  a 
non-Abelian kind, in which case we must better  understand the quantum mechanical properties of these degrees of freedom.   We would like to understand how the 't Hooft-Polyakov monopoles \cite{TH}   (arising from a gauge symmetry breaking, $G \to H$)  and Abrikosov-Nielsen-Olesen vortices \cite{ANO}  (of a broken gauge theory 
$H \to \mathbbm 1$)   are generalized in situations where the relevant gauge group $H$ is non-Abelian.

The key developments which allowed us a qualitatively better understanding of these solitons are the following.  First, the Seiberg-Witten solutions of  ${\cal N} =2$  supersymmetric gauge theories \cite{SW1,SW2} revealed the quantum-mechanical behavior of the magnetic monopoles in an unprecedented fashion. In the presence of  matter fields (quarks and squarks) these theories have, typically,  vacua with non-Abelian dual gauge symmetry in the infrared \cite{APS,CKM}.  Thus in these systems non-Abelian monopoles do exist and play a central role in confinement and dynamical symmetry breaking.  Second, the discovery of non-Abelian vortex solutions \cite{HT,ABEKY}, i.e., soliton vortices with continuous, non-Abelian moduli, has triggered an intense research activity on the classical and quantum properties of these solitons, leading to a rich variety of new interesting results \cite{TongTasi, SYReview, EtoetalModuli}.

\section{Monopoles}

When the gauge symmetry is spontaneously broken
\be
  G   \,\,\,{\stackrel {\brc \phi_{1} \ckt    \ne 0} {\longrightarrow}} \,\,\, H   \label{this}
\ee
where $H$ is some non-Abelian subgroup of $G$,   the system possesses  a set of regular magnetic monopole
solutions  in the  semi-classical approximation.  They are  natural generalizations  of the Abelian   't Hooft-Polyakov monopoles \cite{TH},
 found originally  in the $G=SO(3)$ theory broken to  $H=U(1)$ by a Higgs mechanism.   
 The gauge field looks asymptotically as
 \begin{equation}   F_{ij} =  \epsilon_{ijk} B_k = 
\epsilon_{ijk}  \frac{ r_k 
}{      r^3}  ({ \beta} \cdot  {\bf H}),           \end{equation}
in an appropriate gauge, where ${\bf H}$ are the diagonal generators of $H$ in the Cartan subalgebra. 
A straightforward generalization of the Dirac's quantization condition leads to \cite{GNO}
\begin{equation}  2 \, {\beta \cdot \alpha} \in  { \bf Z}   \label{naqcond}
\end{equation}
where $\alpha$ are the root vectors of $H$. In the simplest such case, $G=SU(3)$, $H= SU(2)\times U(1)/{\mathbbm Z}_{2}\sim U(2)$, a straightforward
idea that the degenerate monopole solutions to be a doublet of the unbroken $SU(2)$  leads however to the well-known difficulties \cite{CDyons,DFHK}. 

On the other hand, the quantization condition Eq.~(\ref{naqcond}) implies that the monopoles should transform, if any, under the dual of $U(2)$:  the individual solutions are labelled by $\beta$ which live in the weight vector space of  ${\tilde H}$,  generated by the  dual roots, 
\be   \alpha^{*} = \frac{\alpha}{\alpha \cdot \alpha}. 
\label{dualg}\ee
As transformation groups of fields $H$ and $\tilde H$ are relatively non-local, the sought-for transformations of monopoles must look as non-local field transformations from the point of view of the original theory \cite{Duality}. 

But the most significant fact is that fully quantum mechanical monopoles appears in the low-energy dual description of a wide class of ${\cal N}=2$ supersymmetric QCD \cite{APS,CKM}.  There must be ways to understand the physics of non-Abelian monopoles starting from a more familiar, semiclassical soliton 
picture.

\section{Vortices}  

Attempts to understand the semi-classical origin of the non-Abelian monopoles appearing in the so-called $r$-vacua of the ${\cal N}=2$  supersymmetric $SU(N)$ 
gauge theory, has eventually led to the discovery of the non-Abelian {\it vortices} \cite{HT,ABEKY}.  They are natural generalizations of the Abrikosov-Nielsen-Olesen (ANO) vortex.  Unlike the ANO vortex, however, the non-Abelian vortices carry continuous zeromodes, i.e., it has a nontrivial moduli. 
 
The simplest model in which these vortices appear is an $SU(N) \times U(1)$ gauge theory with $N_{f}=N$ flavors of squarks in the fundamental representation. The secret of the non-Abelian vortices lies in the so-called color-flavor locked phase, in which the squark fields (written as $N \times N$ color-flavor mixed matrix) takes the VEV of the  form, 
\beq   \brc q(x) \ckt =   v\, {\mathbbm 1}_{N\times N}\;.
\eeq
The $SU(N)$ gauge symmetry is completely broken, but the color-flavor mixed diagonal symmetry remains unbroken.   

The vortex configuration in this vacuum involves scalar fields of the form, 
\beq      q(x)  = v\,  \left(\begin{array}{cccc}e^{i \phi} \, f(\rho) & 0 & 0 & 0 \\0 & g(\rho) & 0 & 0 \\0 & 0 & \ddots & 0 \\0 & 0 & 0 & g(\rho)\end{array}\right)
\label{esempio}\eeq
where $\rho, \phi, z$ (the static vortex does not depend on  $z$)  are the cylindrical coordinates.  The gauge fields take appropriate form, in order to ensure that the kinetic term tends to zero asymptotically, ${\cal D} q(x) \to 0$.     In Eq.~(\ref{esempio})  the first flavor of the squark winds, but the full solution
$A_{i}, q(x)$ can be rotated in the color flavor  $SU(N)$  transformations, 
\[   A_{i}, \to   U (A_{i} + i \de_{i}) U^{\dagger},  \qquad  q(x)  \to   U q(x) U^{\dagger}\;, 
\]  
leaving the tension invariant.

In other words, individual vortices break the exact $SU(N)_{C+F}$ symmetry of the system, developing therefore non-Abelian orientational zeromodes. 
Its nature is seen from the fact that the vortex Eq.~(\ref{esempio}) breaks the global symmetry as 
\be   SU(N) \to  SU(N-1) \times U(1)/{\mathbbm Z}_{N-1}; 
\ee 
 the vortex  moduli is given by  
 \be      CP^{N-1} \sim   \frac{ SU(N) }{  SU(N-1) \times U(1)/{\mathbbm Z}_{N-1}}\;.
 \ee
 They are Nambu-Goldstone modes, which however can propagate only inside the vortex: far from it  they are massive. 
 
 The quantum properties of the non-Abelian orientational modes (the effective $CP^{N-1}$  sigma model),  the study of non-Abelian vortices of higher winding numbers, the generalization  to the cases of larger number of flavors and the study of the resulting, much richer vortex moduli spaces, 
the question of vortex stability in the presence of small non-BPS corrections, extension to more general class of gauge theories, etc.  have been the subjects of an intense research activity  recently.

\section{Non-Abelian vortices with general gauge groups}

One of the new results by us \cite{General} is the construction of non-Abelian vortex solutions based on a general gauge group $G^{\prime} \times U(1)$,  where  $G^{\prime}=SU(N),$ $SO(N),$  $USp(2N)$, etc.  As in models based on $SU(N)$ gauge groups studied extensively in the last few years,  we work with simple models which have the structure of the bosonic sector of $N=2$ supersymmetric models.  The model contains a  FI (Fayet-Iliopoulos)-like term in the $U(1)$ sector, allowing the system to develop stable vortices. A crucial aspect  is that we work in a complete Higg vacuum,  but with an unbroken color-flavor diagonal symmetry. 
We take as our model system 
\bqa
\cal {L} & =& \Tr_c  \Big[
-\frac{1}{2e^2}F_{\mu\nu}F^{\mu\nu}
-\frac{1}{2g^2}\hat{F}_{\mu\nu}\hat{F}^{\mu\nu}
+\cal{D}_\mu    H\left(\cal{D}^\mu H\right)^{\dagger}
\nonumber \\  
&& -\frac{e^2}{4}\left|X^0t^0 - 2\xi t^0\right|^2  -\frac{g^2}{4}\left|X^at^a\right|^2 \Big]\ , \label{Lagrang}\nonumber
\eea
with the field strength, gauge fields and covariant derivative denoted as 
\bqa
F_{\mu\nu} &  =&   F_{\mu\nu}^0t^0 \,, \quad 
F_{\mu\nu}^0   = \partial_\mu A_\nu^0 - \partial_\nu A_\mu^0 \ , \quad
\hat{F}_{\mu\nu} = \partial_\mu   A_{\nu} - \partial_\nu A_\mu +
i\left[A_\mu,A_\nu\right],   \nonumber \\
A_\mu &=& A_\mu^at^a \ , \quad 
{\cal D}_{\mu}=  \partial_{\mu}
  +  i  A_{\mu}^{0}   t^{0} +  i A_{\mu}^{a}  t^{a},    
\nonumber  \eea
$A_\mu^0$ is the gauge field associated with $U(1)$
and $A_\mu^a$ are the gauge fields of $G'$. 
The matter scalar fields are written 
as an $N \times N_{\rm F}$ complex 
color (vertical) -- flavor (horizontal) mixed matrix $H$. 
It can be expanded as
$ X = HH^\dag = X^0t^0 + X^at^a + X^\alpha t^\alpha \ \,\, $, 
$X^0 = 2\,\Tr_c\left(HH^\dag t^0\right)$, 
$X^a = 2\,\Tr_c\left(HH^\dag t^a\right)$. 
$t^0$ and $t^a$ stand for the 
$U(1)$ and $G'$ generators, respectively, 
and finally,  
$t^\alpha \in \mathfrak{g}'_{\perp}$, where
$\mathfrak{g}'_{\perp}$ is the orthogonal complement 
of the Lie algebra $\mathfrak{g}'$ in $\mathfrak{su}(N)$. 
The traces with subscript $c$ are over the color indices. 
$e$ and $g$ are the $U(1)$ and $G'$ coupling constants, respectively,
while $\xi$ is a {real} constant.

We choose the maximally  ``color-flavor-locked'' vacuum of the system, 
\beq
\langle H \rangle = \frac{v}{\sqrt{N}}{\bf 1}_{N} \ ,
\qquad \xi = \frac{v^2}{\sqrt{2N}}\,. \label{eq:vacuum} 
\eeq
We have taken  $N_{\rm F}=N$  which is the minimal number of flavors allowing for such a vacuum.  Note that, unlike the $U(N)$  model studied extensively in the last several years,  the vacuum is not unique in these cases (i.e., with a general gauge group), even with such a minimum choice of the flavor multiplicity. This difference may be traced to  the fact that groups such as $SO(N)\times U(1)$ and $USp(N)\times U(1)$ form strictly smaller subgroups of $U(N)$. 

    The existence of a continuous vacuum degeneracy implies the emergence of vortices of semi-local type; this aspect will be crucial in the discussion of the fractional vortices in the second part of this talk.  However, for now, we stick to the particular vacuum  Eq.~(\ref{eq:vacuum}) and consider vortices and their moduli in this theory.  The standard  Bogomol'nyi completion reads
    \bqa
    T &=& \int d^2x \,\Tr_c   \Big[
\frac{1}{e^2}\left|F_{12}-\frac{e^2}{2}\left(X^0t^0 -
2\xi t^0\right)\right|^2
+\frac{1}{g^2}\left|\hat{F}_{12}-\frac{g^2}{2}\,X^at^a\right|^2  \nonumber \\
& &  +4\left|\bar{\cal{D}}H\right|^2-2\xi F_{12}t^0 \Big]
\ge -\xi \int d^2x\, F^0_{12} \ ,
\eea
where $\bar{\cal  D } \equiv
\frac{{\cal D}_1+i {\cal D}_2}  {2}$,  
 $z = x^1+ix^2$.  In the BPS limit one has
\beq T = 2\sqrt{2N}\pi\xi\nu = 2\pi v^2 \nu\ , \qquad
\nu = -\frac{1}{2\pi\sqrt{2N}}\int d^2x\,F_{12}^0
\ ,  \label{eq:tension} \eeq
where  $\nu$ is the $U(1)$ winding number of the vortex.
This leads immediately to the  vortex BPS equations
\bqa
\bar{\cal{D}}H &=& \bar{\partial}H + i\bar{A}H = 0 \ , \label{BPS1} \\
F_{12}^0 &=&   e^2\left[\, \Tr_c\left(HH^\dag t^0\right) - \xi \, \right] \ ,
\qquad 
F_{12}^a   = g^2\, \Tr_c\left(HH^\dag t^a\right) \ . \label{BPS3}
\eea
 The matter BPS equation (\ref{BPS1}) can be solved  
by the Ansatz
\beq 
H = S^{-1}(z,\bar{z})H_0(z) \ , \qquad 
 \bar{A} = -iS^{-1}(z,\bar{z})\bar{\partial}S(z,\bar{z}) \ , 
\label{eq:H_A}
\eeq
where $S$ belongs to the complexification of the gauge
group,  $S\in \mathbb{C}^{*}\times {G'}^\mathbb{C}$. 
$H_0(z)$, holomorphic in $z$, is    the \emph{moduli matrix},  which contains all moduli parameters of the
vortices. 

A gauge invariant object can be constructed from $S$    as $\Omega = SS^\dag$. This can be conveniently split into the $U(1)$ part and
the $G'$ part, so that $S = s\, S'$ and analogously 
$\Omega = \omega\, \Omega'$, $\omega  = |s|^2$, $\Omega' =S'{S'}^\dag$.   
The tension (\ref{eq:tension}) can be rewritten
as  
\be
T = 2\pi v^2\nu = 2v^2\int d^2x\  \partial\bar{\partial}\log\omega  \ , \qquad
\nu = \frac{1}{\pi}\int d^2x\, \partial\bar{\partial}\log\omega \ ,
\ee
and  $\nu$ determines  the asymptotic behavior of the Abelian field 
as
\[ 
\omega = ss^\dag \sim \left|z\right|^{2\nu}, \hspace{10 mm} \mbox{for}\ 
  \left|z\right|\to\infty\ . 
\] 

The minimal vortex solutions can then be written down by
making use of the holomorphic invariants for the gauge group $G'$ made
of $H$, which we denote as  $I_{G'}^i(H)$. If the $U(1)$ charge of the
$i$-th invariant is  $n_i$, $I_{G'}^i(H) $ satisfies     
\[
I_{G'}^i(H) = I_{G'}^i\left(s^{-1}{S'}^{-1}H_0\right) = 
s^{-n_i}I_{G'}^i(H_0(z)) \ , 
\]
while the boundary condition is
$
I_{G'}^i(H)\bigg|_{|z|\to\infty}=I_{\rm vev}^i  \, e^{i\nu n_i\theta}
\ ,  
$
where $\nu \, n_i$ is the number of the {\it zeros} of $I_{G'}^i$. 
This leads then to the following asymptotic behavior
\[ 
I_{G'}^i(H_0) = s^{n_i}I_{G'}^i(H) 
   \,\,\,{\stackrel {|z|\to\infty} {\longrightarrow}} \,\,\,
 I_{\rm   vev}^i z^{\nu n_i} \,. 
\]
It shows that $I_{G'}^i(H_0(z))$,  being holomorphic in  $z$,  are
actually polynomials. Therefore $\nu \, n_i$ must be positive integers
for all $i$:
\[ 
\nu \, n_i \in \mathbb{Z}_{+} \qquad \to \qquad \nu = \frac{k}{n_0}
\ , \qquad k\in\mathbb{Z}_{+} \ , \label{defnu} 
\]
with 
$
n_0 \equiv {\rm gcd}\left\{n_i\,|I_{\rm vev}^i\neq 0\right\}.
$
 The $U(1)$
gauge transformation $e^{2\pi i /n_0}$  leaves $I_{G'}^i(H)$ invariant
and thus the true gauge group is 
\[ 
G = \left[U(1)\times G'\right]/\mathbb{Z}_{n_0} \ , 
\]
where $\mathbb{Z}_{n_0}$ is the center of the group $G'$.  
The minimal winding in $U(1)$ found here,  $\frac{1}{n_0}$,
corresponds to the minimal element of  
$\pi_{1}(G) = {\mathbb Z}$:   it represents a minimal loop in our
group manifold $G$.  
As a result we find the following non-trivial constraints for $H_0$ 
\[ 
I_{G'}^i(H_0) = I_{\rm vev}^i \, z^{\frac{k n_i}{n_0}} +
{O}  \left(z^{\frac{k n_i}{n_0} -1}\right) \ . 
\] 

\subsection{GNO quantization for non-Abelian vortices}\label{sec:special_mm}

 Certain special solutions of a given theory  can
be found readily, as follows.   It turns out that each such solution is characterized by a 
{\it weight vector of the dual group}, and 
is  parametrized  by a set of integers $\nu_a$
$\left(a=1,\cdots,{\rm rank}(G') \right)$
\beq
H_0 (z) = z^{\nu {\bf 1}_N + \nu_a {\cal H}_a}
\in U(1)^{\mathbb C} \times {G'}^{\mathbb C}\ ,
\eeq
where $\nu=k/n_0$ is the $U(1)$ winding number and ${\cal H}_a$
are the generators of the Cartan subalgebra of $\mathfrak{g}'$. 
$H_0$ must be holomorphic in $z$ and {\it single-valued}, which gives
the constraints for a set of integers $\nu_a$
\[
\left( \nu {\bf 1}_N + \nu_a {\cal H}_a \right)_{ll} 
\in {\mathbb Z}_{\ge 0} \quad \forall \,l\ .
\label{eq:quant1}
\]
Suppose that we now consider scalar fields in an $r$-representation of
$G'$. The constraint is equivalent to
\beq
\nu + \nu_a \mu_a^{(i)}\in {\mathbb Z}_{\ge 0} \quad \forall \,i\ ,
\label{eq:quant2}
\eeq
where 
$\vec \mu^{(i)}=\mu_a^{(i)}$ $\left(i=1,2,\cdots,{\rm dim}(r)\right)$ 
are the weight vectors for the $r$-representation of $G'$.
Subtracting pairs of adjacent weight vectors, one arrives at the
quantization condition 
\beq
\vec \nu \cdot \vec \alpha \in {\mathbb Z}\ ,  \label{GNOWQ}
\eeq
for every {\it root vector}  $\alpha$  of $G'$.

Now  Eq.~(\ref{GNOWQ}) is formally identical to the well-known
GNO {\it monopole} quantization condition \cite{GNO}, as well as to the na\"{i}ve vortex flux quantization rule \cite{KS}. There is however a crucial difference here,  from these earlier results. 
Because of an exact flavor (color-flavor diagonal $G_{\rm C+F}$)
symmetry, broken by individual vortex
solutions, our vortices possess continuous orientational moduli. These zero modes are normalizable, 
unlike those encountered in the earlier attempts to define quantum ``non-Abelian  monopoles''.  


These non-Abelian modes of our vortices---they are a kind of  Nambu-Goldstone modes---can fluctuate and propagate along the vortex length.  In systems with a  hierarchical symmetry breaking,  
 \[    G_{0} \to     G=G^{\prime} \times U(1)   \to {\mathbf 1},
\]
 where our  $G= G^{\prime}\times U(1)$  model might emerge as the low-energy approximation, 
these orientational zero modes   get absorbed by massive monopoles
at the vortex extremities.  This process endows the monopoles with fully quantum-mechanical  non-Abelian (GNO-dual)  charges, 
as has been suggested  by the author and others in several occasions \cite{Duality},   but we shall not dwell on this subject further here.

The solution of the quantization condition (\ref{GNOWQ}) is that 
\[
 \tilde{\vec\mu} \equiv \vec\nu/2 \ ,
\]
is any of the {\it weight vectors} of the dual group of $G^{\prime}$. 
The dual group, denoted as $\tilde G'$, is defined by the dual root 
vectors \cite{GNO} 
$
\vec \alpha^* = {\vec\alpha}/({\vec\alpha \cdot \vec\alpha})\ .
$
Examples of  dual pairs of groups $G'$, $\tilde G'$, are shown  in Table
\ref{tabledualgroup}. 
\begin{table}[tb]
\begin{center}
\begin{tabular}{c|c}
$G'$ & $\tilde G'$  \\
\hline
\hline 
$SU(N)$  &  $SU(N)/{\mathbb Z}_{N}$ \\
 $U(N)$  &  $ U(N)$ \\
$SO(2M)$ & $SO(2M)$ \\
$USp(2M) $ & $SO(2M+1)$ \\
$SO(2M+1)$ & $USp(2M)$ 
\end{tabular}
\caption{Some pairs of dual groups}
 \label{tabledualgroup}
\end{center}
\end{table}
Note that (\ref{eq:quant2}) is actually  stronger than (\ref{GNOWQ}), the l.h.s. must be a
nonnegative  integer. This positive quantization condition allows for a few weight vectors only. 
For concreteness, let us consider scalar fields in the fundamental
representation, and choose a basis where the Cartan generators of 
$G'=SO(2M),SO(2M+1),USp(2M)$ are given by 
\beq
{\cal H}_a = {\rm diag}
\Big(\underbrace{0,\cdots,0}_{a-1},\frac{1}{2},
\underbrace{0,\cdots,0}_{M-1},-\frac{1}{2},0,\cdots,0\Big)\ ,
\eeq
with $a=1,\cdots,M$. In this basis, special solutions $H_0$ have the
form   for $G'=SO(2M)$ and $USp(2M)$
\beq
H_0^{\left( \tilde{\mu}_1,\cdots, \tilde{\mu}_M \right)} = {\rm diag}\left(
z^{k_1^+},\cdots,z^{k_M^+},z^{k_1^-},\cdots,z^{k_M^-}
\right)\ ,
\label{eq:special_2M}
\eeq
while for $SO(2M+1)$
\beq 
H_0^{\left( \tilde{\mu}_1,\cdots, \tilde{\mu}_M \right)} = {\rm diag}
\left(
z^{k_1^+},\cdots,z^{k_M^+},z^{k_1^-},\cdots,z^{k_M^-},z^k
\right)\ ,
\label{eq:sp_points_so_odd}
\eeq
where $k_a^{\pm}=\nu \pm \tilde{\mu}_a$. 

For example, in the cases of $G'=SO(4),USp(4)$ with a $\nu=1/2$
vortex, there are four special solutions with 
$\vec{\tilde{\mu}} = (\frac{1}{2},\frac{1}{2}),(\frac{1}{2},-\frac{1}{2}),
(-\frac{1}{2},\frac{1}{2}),(-\frac{1}{2},-\frac{1}{2})$
\bqa
H_0^{(\frac{1}{2},\frac{1}{2})} &=& 
{\rm diag}(z,z,1,1) = z^{\frac{1}{2}{\bf 1}_4 + 1\cdot{\cal H}_1 + 1\cdot{\cal H}_2},\\
H_0^{(\frac{1}{2},-\frac{1}{2})} &=& 
{\rm diag}(z,1,1,z) = z^{\frac{1}{2}{\bf 1}_4 + 1\cdot{\cal H}_1 - 1\cdot{\cal H}_2},\\
H_0^{(-\frac{1}{2},\frac{1}{2})} &=& 
{\rm diag}(1,z,z,1) = z^{\frac{1}{2}{\bf 1}_4 - 1\cdot{\cal H}_1 + 1\cdot{\cal H}_2},\\
H_0^{(-\frac{1}{2},-\frac{1}{2})} &=& 
{\rm diag}(1,1,z,z) = z^{\frac{1}{2}{\bf 1}_4 - 1\cdot{\cal H}_1 - 1\cdot{\cal H}_2}.
\eea
These four vectors are the same as the weight vectors of two Weyl spinor
representations 
${\bf 2} \oplus {\bf 2}'$
of $\tilde G'=SO(4)$ for $G'=SO(4)$,
and the same as those of the Dirac spinor representation
${\bf 4}$ of $\tilde G'=Spin(5)$ for $G'=USp(4)$. 

\begin{figure}[htb]
\begin{center}
\includegraphics[height=1.4cm]{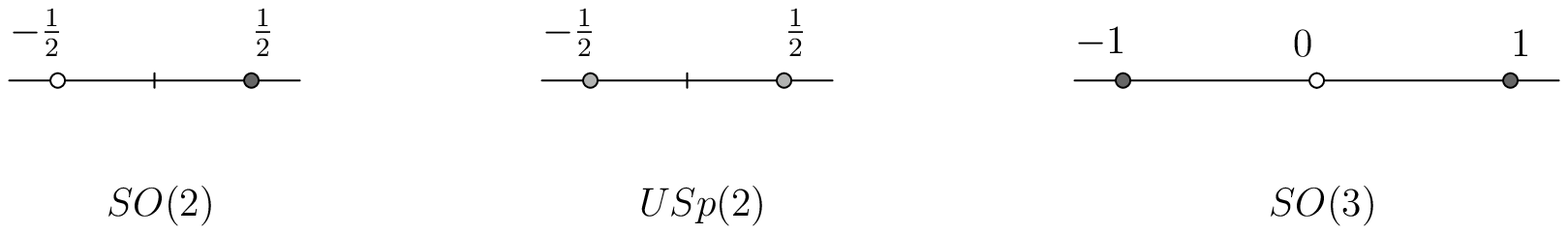}\\\ \\
\includegraphics[height=3.8cm]{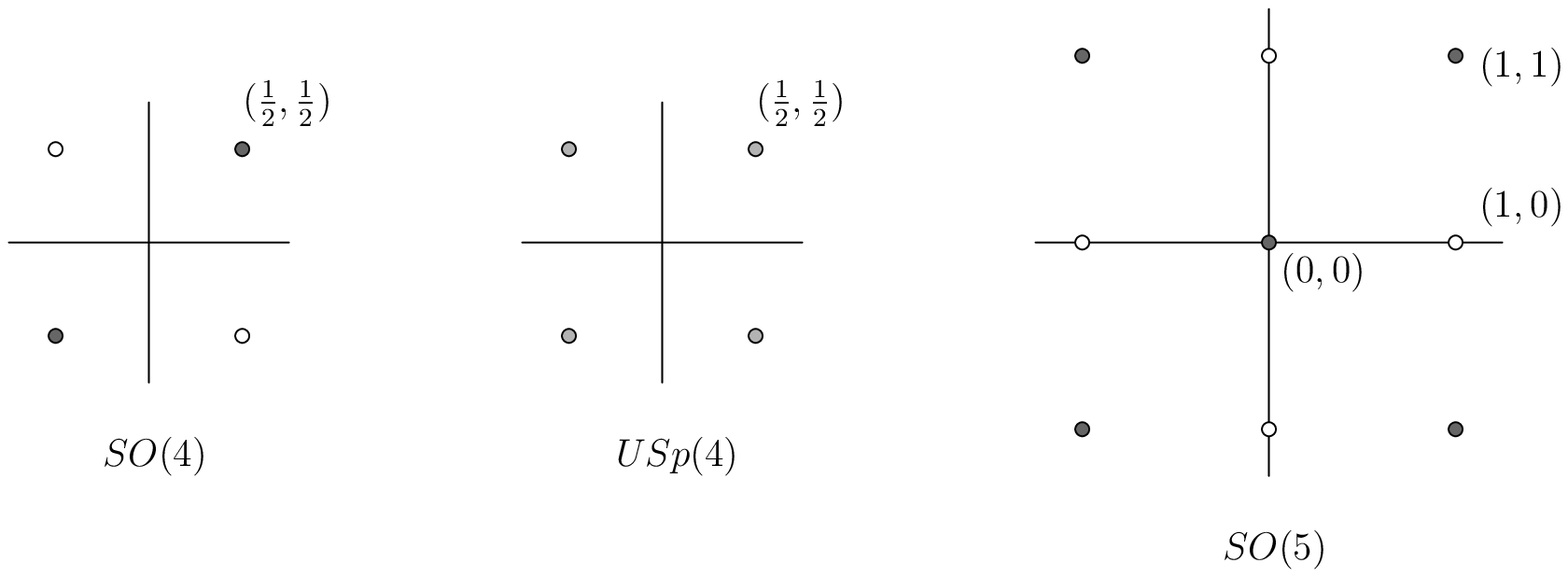}\\\ \\
\includegraphics[height=3.2cm]{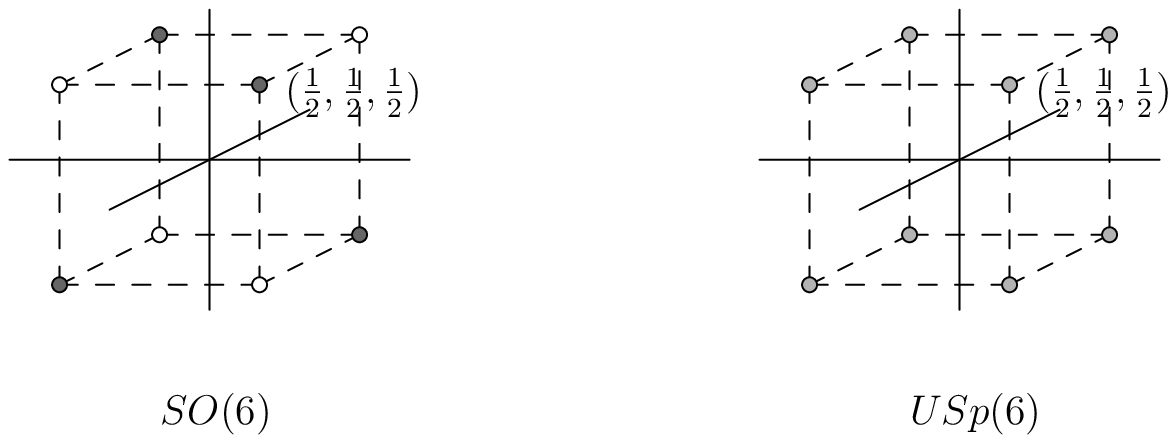}
\caption{The special points for the $k=1$ vortex.}
\label{fig:patches_k1}
\end{center}
\end{figure}

The weight vectors corresponding to the $k=1$ vortex 
in various gauge groups are shown in Fig.~\ref{fig:patches_k1}.  In all cases the results found are consistent with the GNO duality. 

\section{Fractional vortices and lumps}

Another exciting recent result concerns the fractional vortex and lumps.\cite{Fractional} 
We have pointed out above that in a general class of gauge theories the  vacuum is not unique, even if the Fayet-Iliopoulos term is present and even if the number of the flavors is the minimum possible for a ``color-flavor'' locked vacuum to exist. In other words, there is a nontrivial vacuum degeneracy, or the {\it vacuum moduli}  ${\cal M}$.   In the first part of the talk, we were mainly interested in the {\it vortex moduli}  ${\cal V}$, in a particular,  maximally color-flavor locked vacuum.
Here we are going to consider all possible vortices---the vortex moduli ${\cal V}$---on all possible points of the vacuum moduli ${\cal M}$ at the same time.

\begin{figure}
\begin{center}
\includegraphics[width=2.3in]{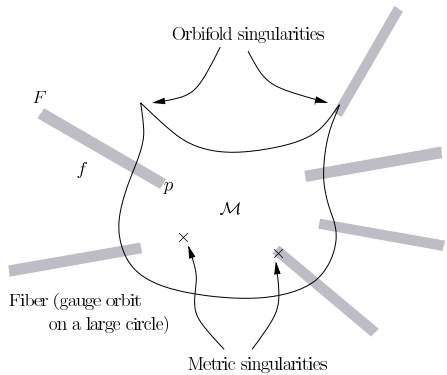}
\caption{\footnotesize Vacuum moduli ${\cal M}$,  fiber $F$ over it,
  and possible singularities} 
\label{Gatto}
\end{center}
\end{figure}

There are in fact two crucial ingredients for the fractional vortex: the  {\it vacuum degeneracy} 
and the {\it BPS saturated}  nature of the vortices.
The first point was emphasized just above: 
the situation is schematically illustrated in Fig.~\ref{Gatto}.
Even if we restrict ourselves to the minimally winding vortex
solutions only,  the vortices represent non-trivial fiber bundles over the vacuum moduli
${\cal M}$.

The BPS-saturated nature of the vortices, on the other hand,  implies that the
vortex equations are reduced to the first-order equations. The
matter equations of motion   are solved by the moduli-matrix Ansatz.   The other equations--the
gauge field equations--reduce, in the strong coupling limit or,
anyway,  sufficiently far from the vortex center, to the vacuum
equations for the scalar fields.  In other words, the vortex
solutions  approximate the sigma model lumps. 
   
\subsection{Structures of the vacuum moduli \label{generalst}} 

\begin{figure}
\begin{center}
\includegraphics[width=2in]{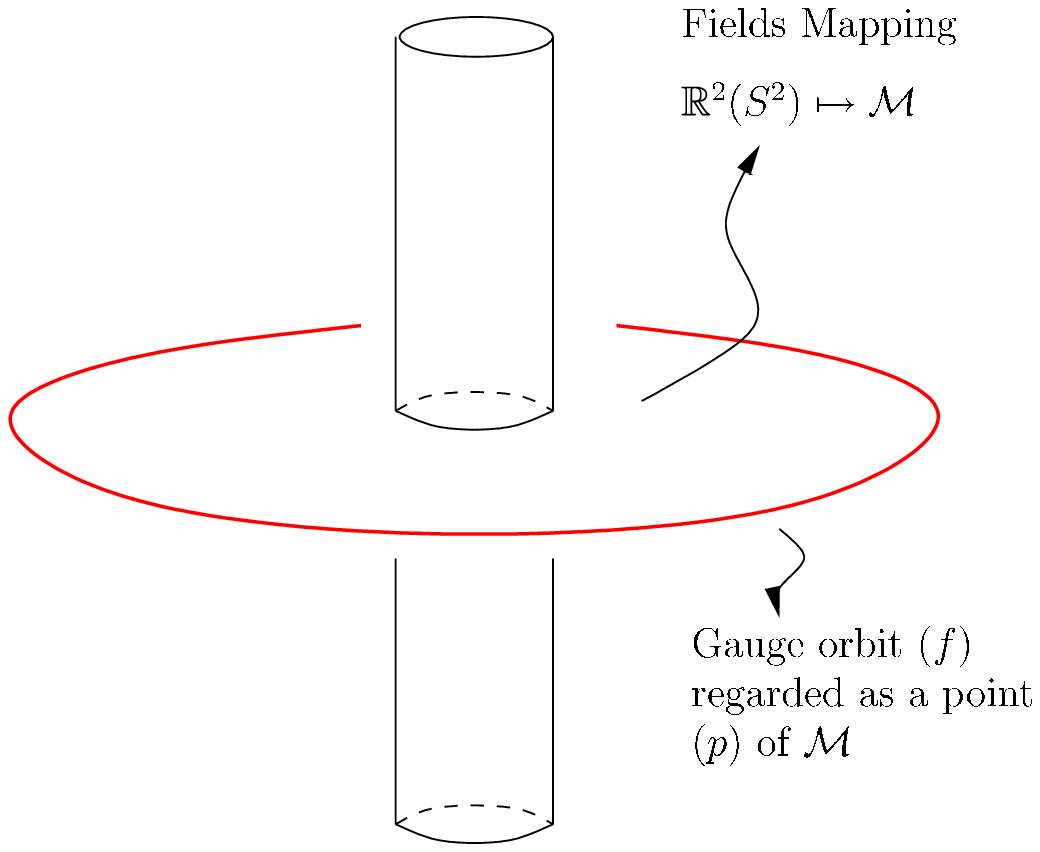}
\caption{ }
\label{VortexFig}
\end{center}
\end{figure}

 Let $M$ be the manifold of the minima of the scalar
potential, the vacuum configuration $M=\{q_{i} \ | \ q^\dag T^I
q = \xi^I\}$.   The vacuum
moduli ${\cal M}$ is given by the points  
\beq  p \in  {\cal M}=  M / F \ , \eeq
where the fiber $F$ is the sum of the gauge orbits of a point in $M$.
A vortex solution is defined on each point of ${\cal M}$, in the sense
that the scalar configuration along a sufficiently large circle
($S^{1}$) surrounding it traces a non-trivial closed orbit in the fiber  $F$ (hence a
point in ${\cal M}$).  The existence of a vortex solution requires
that 
\beq \pi_{1}\left(F, f\right) \neq \mathbf{1} \ ,\eeq
where $f$ is a point in $M$.  The
field configuration on a disk $D^{2}$ encircled by $S^{1}$ traces
${\cal M}$, apart from points at finite radius where it goes off $M$
(hence from ${\cal M}$).  In other words it represents an 
element of $\pi_{2}({\cal M}, p)$, where $p$ is the gauge orbit
containing $f$, or 
$ p = \pi(f) $: 
 $\pi$ is the projection of the fiber onto a point of the base space 
${\cal M}$.   The exact sequence of homotopy groups for the fiber bundle reads 
\[ \cdots \to 
\pi_{2}\left(M, f\right) \to 
\pi_{2}\left({\cal M}, p\right) \to
\pi_{1}\left(F, f\right) \to 
\pi_{1}\left(M, f\right) \to 
\pi_{1}\left({\cal M}, p\right) \to
\cdots   \label{homotseq} \] 
where $\pi_{2}(M/F, f) \sim\pi_{2}({\cal M}, p)$.  
See Fig.~\ref{VortexFig}. 

Given the points $f, p$ and the space ${\cal M}$, the vortex solution
is still not unique.  Any exact symmetry of the system  broken by an individual vortex
solution gives rise to vortex zero modes (moduli), $\cal{V}$.
Our main interest here  however  is the vortex moduli which arises
from the non-trivial vacuum moduli ${\cal M}$ itself.     Due to the BPS
nature of our vortices, the gauge field equation
\[ F_{12}^{I} = g_{I}^{2}\left(q^{\dag} T^{I} q - \xi^I\right) \ ,  \]
reduces,  in the strong-coupling limit (or in any case, sufficiently
far from the vortex center),  to the vacuum equation
defining $M$. 
This means that a vortex configuration can be approximately seen  as a
non-linear $\sigma$-model (NL$\sigma$M) lump with target space ${\cal
  M}$, as was already anticipated.  Various
distinct maps 
$ S^{2}  \mapsto {\cal M}$   
of the same homotopy class correspond to physically inequivalent  
solutions;  each of  these corresponds to a vortex  with the equal
tension  
\[ T_{\rm min}=  -\xi^{I} \int  d^{2}x \, F_{12}^{I} > 0 \ , \]
by their BPS nature. 
They represent non-trivial {\it vortex moduli}.

The semi-local vortices of the so-called extended-Abelian Higgs (EAH) model
 arise precisely this way.   
In an Abelian Higgs model with $N$ flavors of (scalar) electrons,  
$ M = S^{2N-1}$, 
$F = S^{1}$,  
${\cal M} = S^{2N-1}/S = \mathbb{C}P^{N-1}$, 
and the exact  homotopy sequence 
tells us that $\pi_{2}(\mathbb{C}P^{N-1})$ and
$\pi_{1}(S^{1}) $ are isomorphic: each (i.e.~minimum) vortex solution
corresponds to a minimal $\sigma$-model lump solution.  

In most cases discussed in our paper \cite{Fractional}, however, the base space ${\cal M}$
will be various kinds of {\it singular manifolds}: a manifold with
singularities, unlike in the EAH model.  
The nature of the singularities depends on the system and on the
particular point(s) of ${\cal M}$.  
Our BPS degenerate vortices represent (generalized) fiber bundles
with the singular manifolds ${\cal M}$ as the base space. 

\subsection{Two  mechanisms for fractional vortex--lump \label{causes}}


There are two distinct mechanisms leading to the appearance of a fractional vortex.   The first is related to the presence of  orbifold singularities in ${\cal M}$. 
For example, let us consider a ${\mathbbm Z}_{2}$ point $p_{0}$ such as the one appearing in a simple $U(1)$ model with two scalars, one of which has charge $2$.  At this singularity, both $ \pi_{2}\left({\cal M}, p\right) $ and $
\pi_{1}\left(F, f\right) $  make a discontinuous change.  The minimum element of $\pi_{1}\left(F_{0}, f_{0} \right) $ is half  of that of $\pi_{1}\left(F, f \right)$ defined off the singularity, and similarly for 
$ \pi_{2}\left({\cal M}, p_{0} \right) $  with respect to $ \pi_{2}\left({\cal M}, p \right) $, $p \ne p_{0}$. 
Even though the exact homotopy sequence continues to hold  on and off the orbifold point, the vortex defined near such a point will look like a doubly-wound vortex, with two centers (if the vortex moduli parameters are chosen appropriately).  Analogous multi-peak vortex solution occurs near a ${\mathbbm Z}_{N}$  orbifold point of ${\cal M}$. 

Another cause for the appearance of fractional peaks is simple and very general: a deformed sigma model geometry.  This phenomenon  can be best seen by considering our system in the strong coupling limit.  Even if the base point $p$ is a perfectly generic, regular point of ${\cal M}$, not close to any singularity, 
the field configurations in the transverse plane ($S^{2}$) trace the whole vacuum moduli space ${\cal M}$.    The energy  distribution reflects the nontrivial structure of $\cal{M}$ as
the volume of the target space is mapped into the transverse plane,
$\mathbb{C}$
\[ E = 2\int_{\mathbb{C}} \frac{\partial^2 K}{\partial \phi^I\partial\phi^{\dag\bar{J}}}
\partial\phi^I\bar{\partial}\phi^{\dag\bar{J}} = 2\int_{\mathbb{C}} \bar{\partial}\partial K
\ .\]
The field configuration may hit for instance one of the singularities (conic or not), 
or simply the regions of large scalar curvature.  Such phenomena thus occur very generally if the underlying sigma model has a deformed geometry  \footnote{Basically the same phenomenon was found also by Collie and Tong. \cite{CollieTong}}. 
  At such points the energy density will show a peak, not necessarily at the vortex center.  Even at finite coupling, the vortex tension density will exhibit a similar substructure. 


\subsection{Some models}

\begin{figure}[h!tp]
\begin{center}
\begin{tabular}{ccl}
\includegraphics[width=6cm]{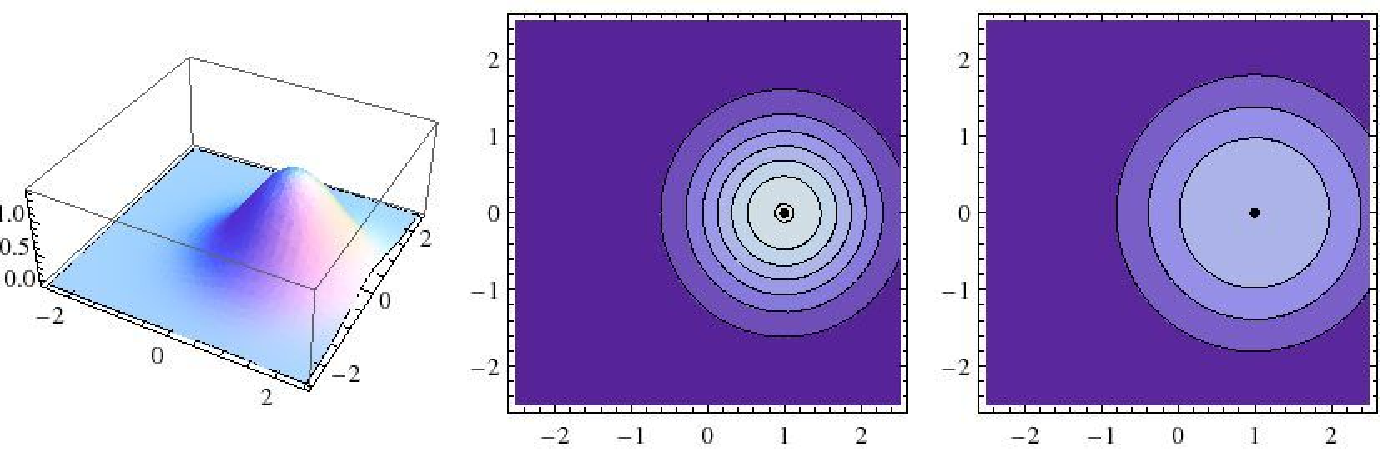} && \includegraphics[height=2cm]{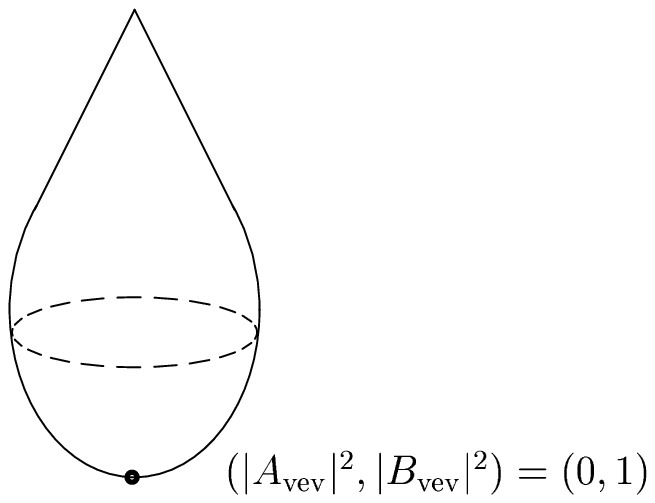}\\
\includegraphics[width=6cm]{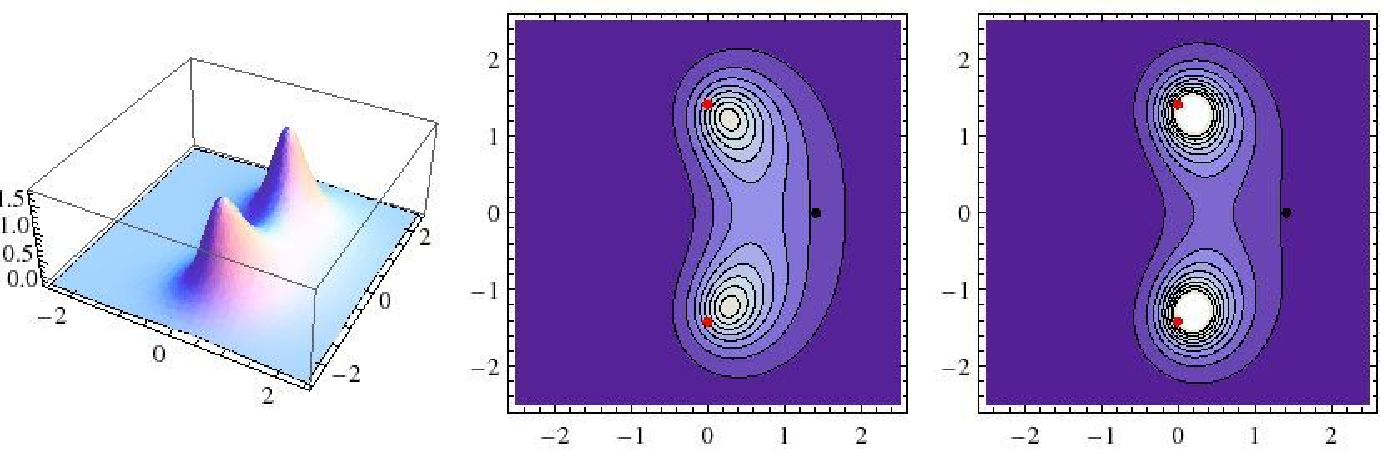} && \includegraphics[height=2cm]{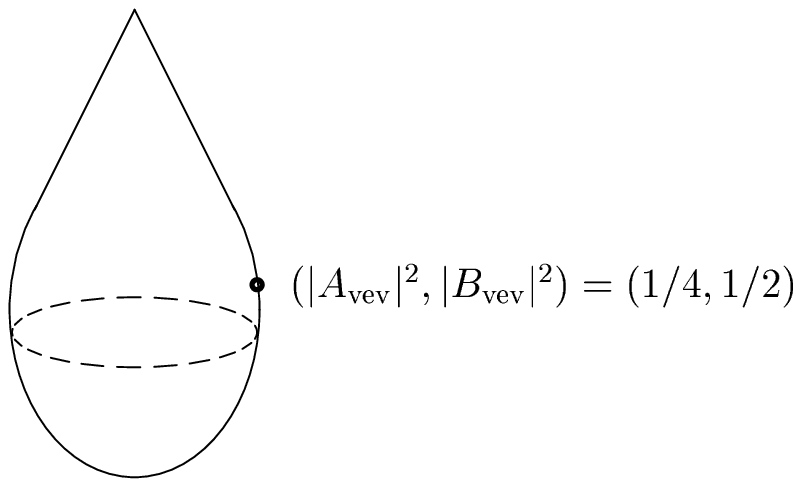}\\
\includegraphics[width=6cm]{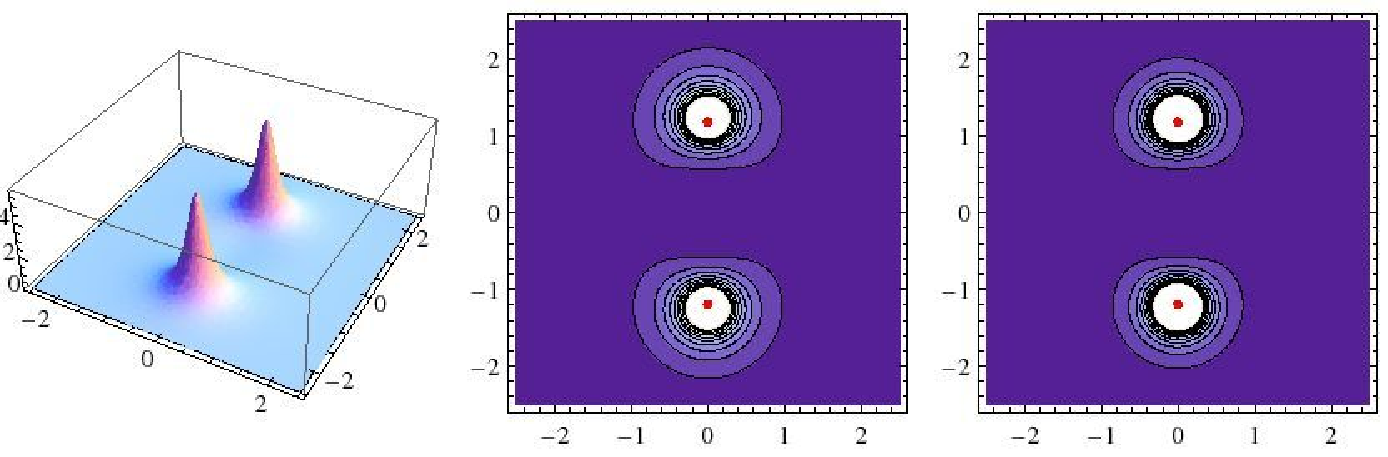} && \includegraphics[height=2cm]{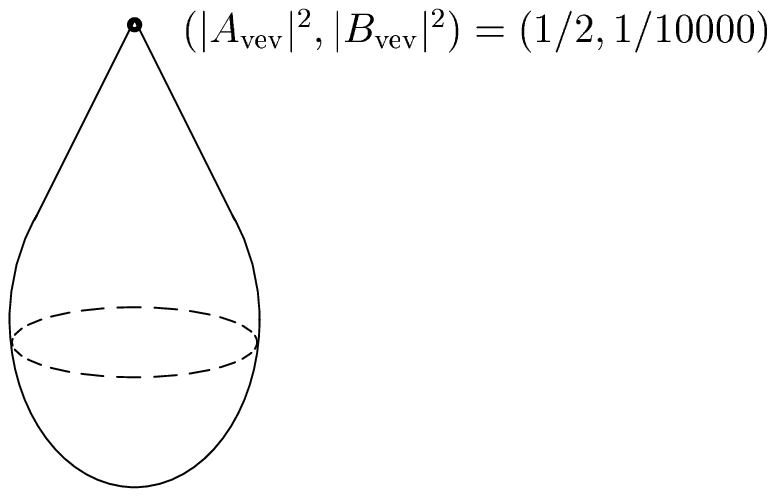} \\
\end{tabular}
\caption{\footnotesize{
  The energy (the left-most and the 2nd left panels) and
 the  magnetic flux  (the 2nd right panels) density, together with
  the boundary values $(A,B)$ (the right-most panel) for the
  minimal vortex. 
  }}
\label{FigMod21}
\end{center}
\end{figure}

A simple model showing the fractional vortex is an extended Abelian Higgs model, with two scalar fields $A$ and $B$ with charges $2$ and $1$, respectively.  Depending on the point of ${\cal M}$ (which is  $\mathbb{C}P^{1}$)  the minimum vortex shows doubly-peaked substructure clearly, see Fig.~\ref{FigMod21}.   The fractional vortex structure in this model nicely illustrates the first mechanism discussed above:   the point $B=0$  is a $Z_{2}$ orbifold point, since there the only nonvanishing field, $A$, having charge $2$, must wind only half of the $U(1)$ to be single-valued.

Another interesting model is a $U_1(1)\times U_2(1)$ gauge theory  with three flavors of scalar
electrons $H = (A,B,C)$ with charges $Q_1 = (2,1,1)$ for 
$U(1)_1$ and $Q_2 = (0,1,-1)$ for $U(1)_2$.  Even though the model has the same $\mathbb{C}P^{1}$ as the vacuum moduli ${\cal M}$ as the first model,  the vortex properties are quite different.  This model turns out to provide a good example of fractional vortex of the second type
(deformed sigma-model geometry). 

\begin{figure}[h!tp]
\begin{center}
\begin{tabular}{ccl}
\includegraphics[width=6cm]{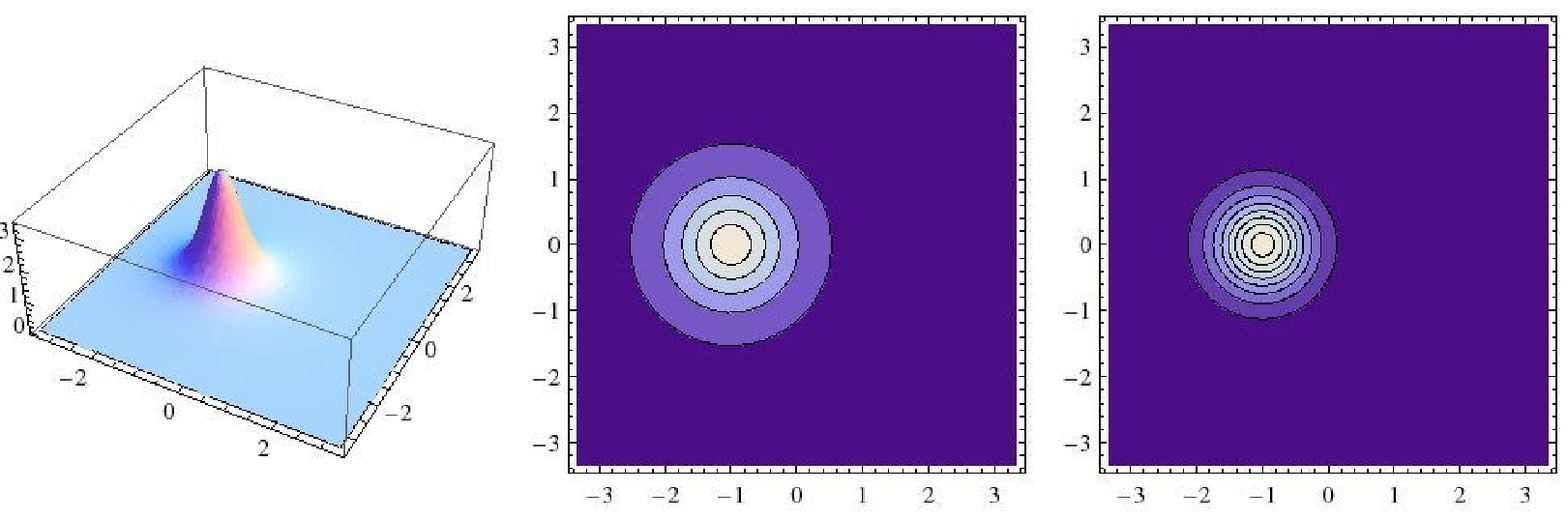} && \includegraphics[height=2cm]{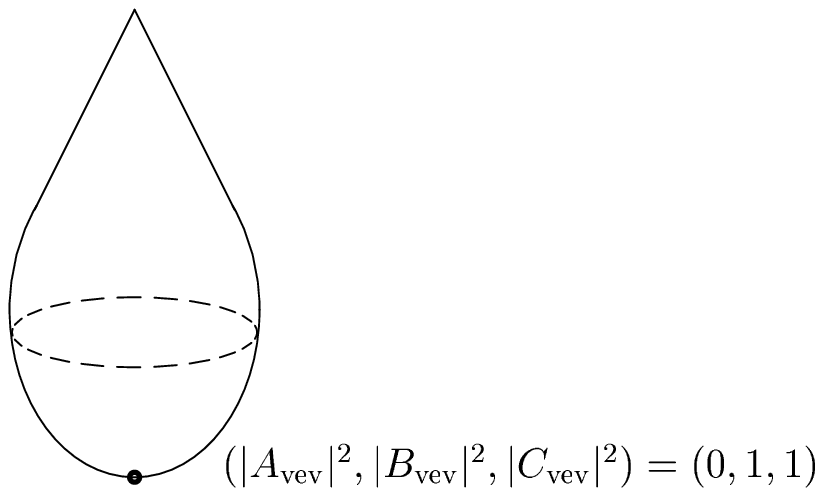}\\
\includegraphics[width=6cm]{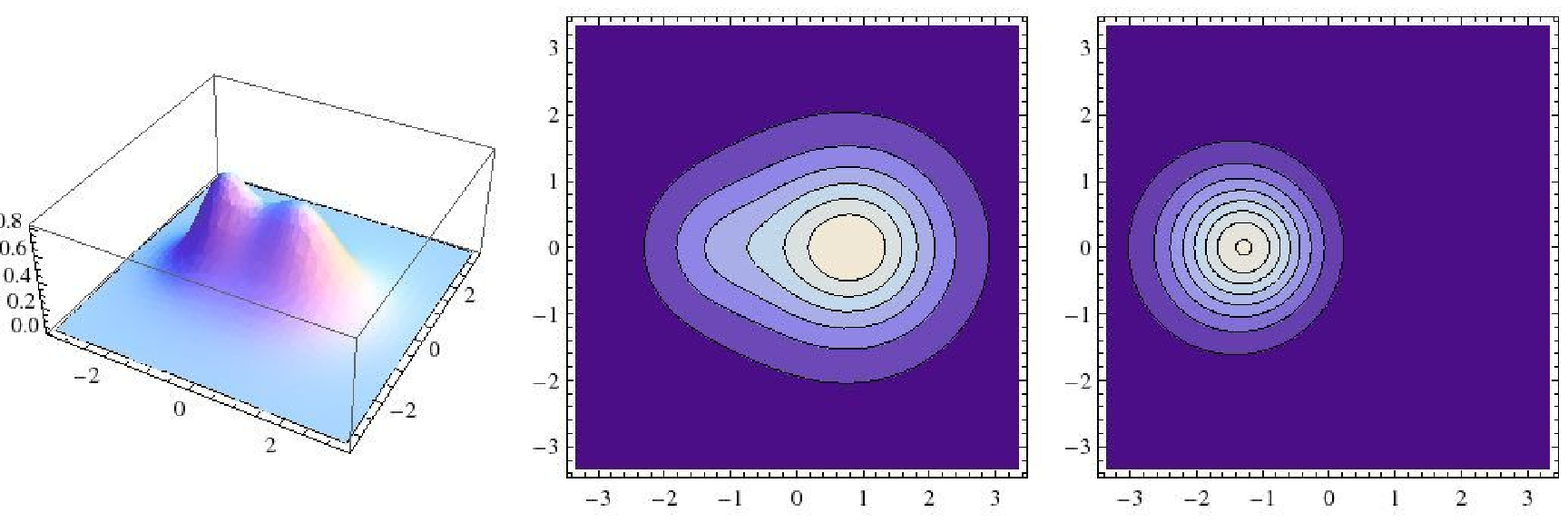} && \includegraphics[height=2cm]{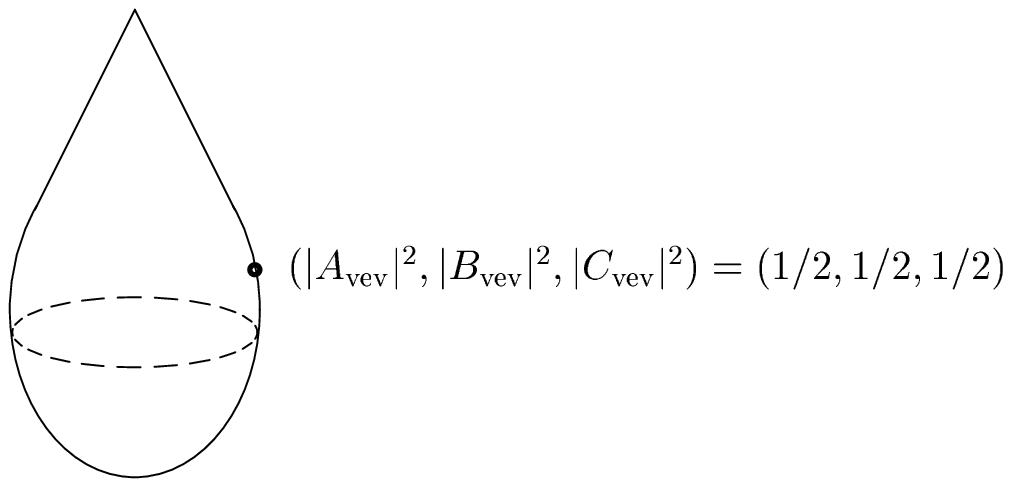}\\
\includegraphics[width=6cm]{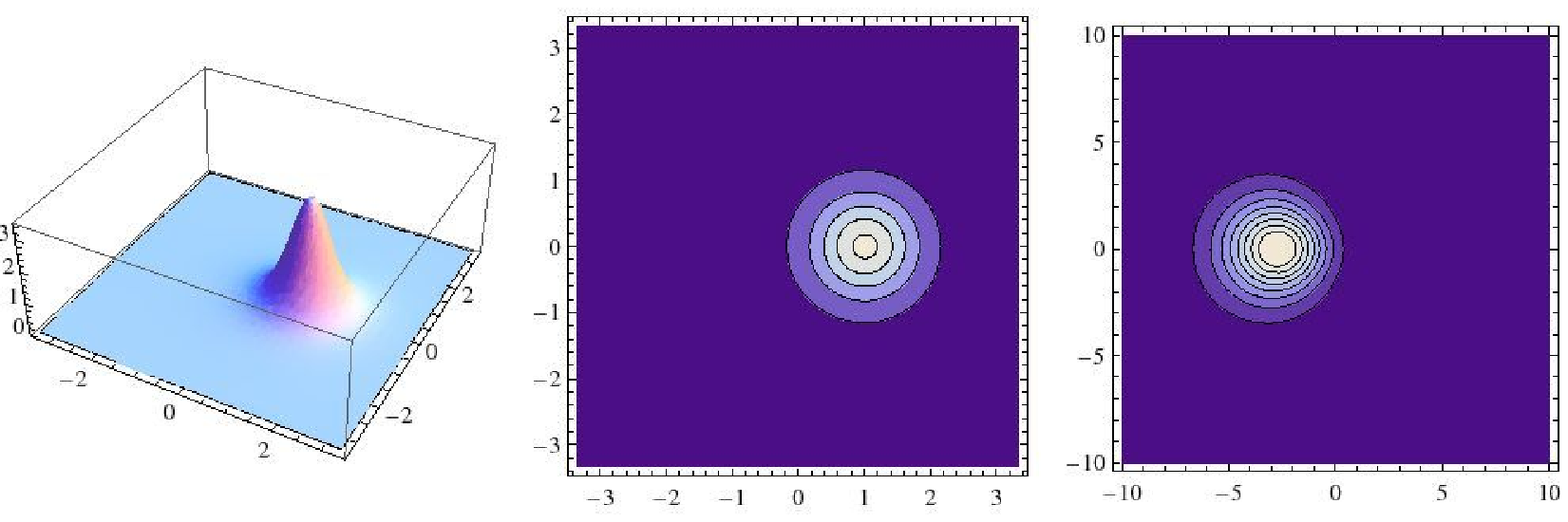} && \includegraphics[height=2cm]{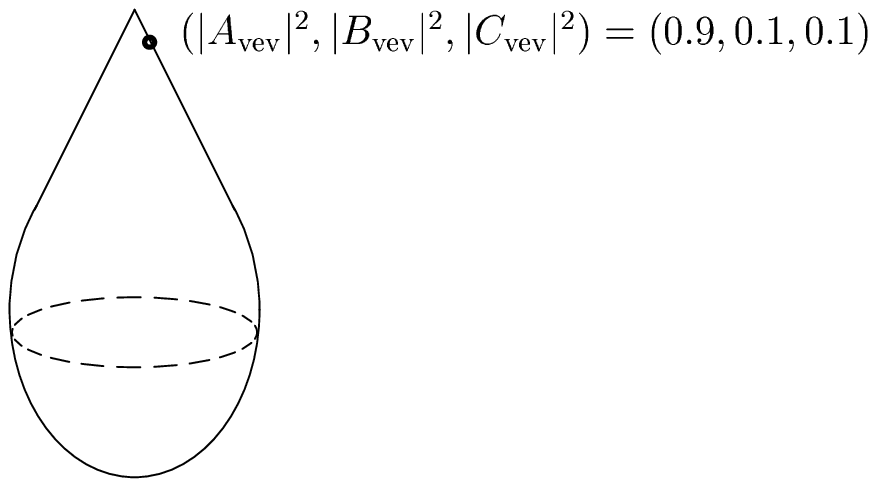}
\end{tabular}
\caption{\footnotesize{The energy density (left-most) and the magnetic
    flux density $F_{12}^{(1)}$ (2nd from the left), $F_{12}^{(1)}$
    (2nd from the right) and the boundary condition (right-most). }}
\label{fig:u1u1_finite}
\end{center}
\end{figure}

\begin{figure}[h!tp]
\begin{center}
\begin{tabular}{cc}
\includegraphics[width=4.5cm]{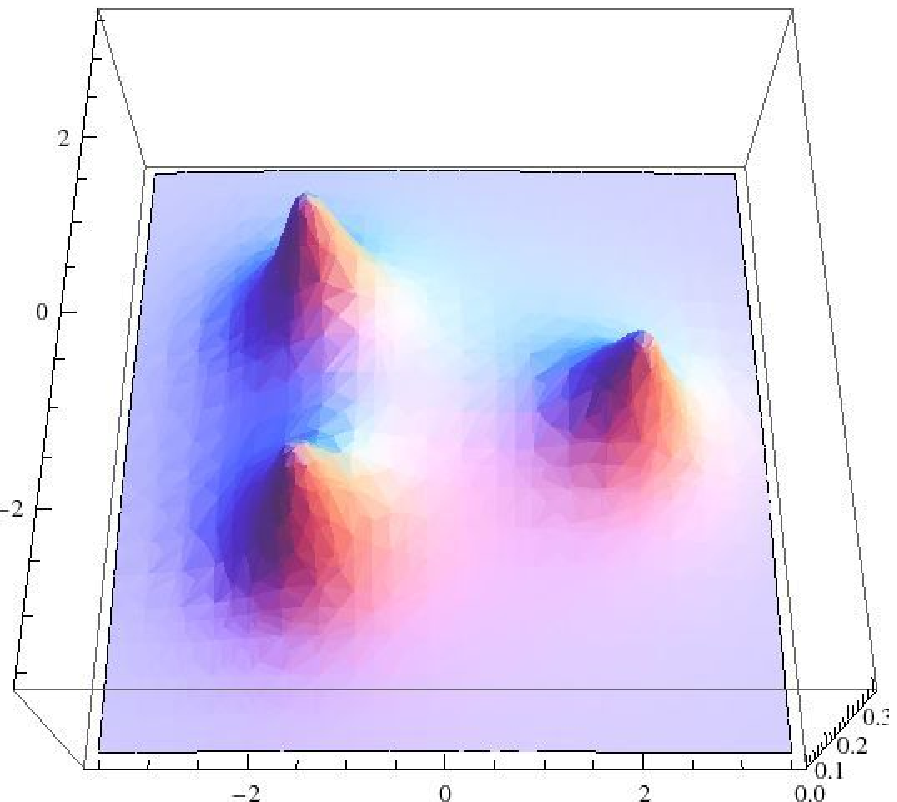} & \includegraphics[width=4cm]{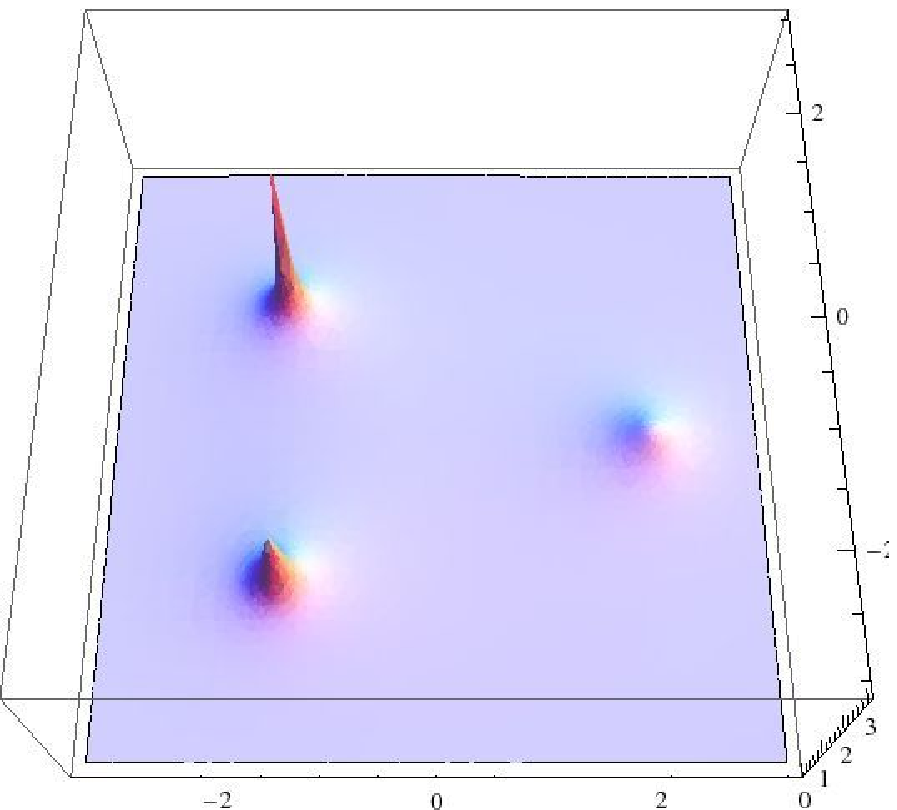}
\end{tabular}
\caption{\footnotesize{
  The energy density of three fractional vortices (lumps) in the
  $U(1)\times SO(6)$ model in the strong coupling approximation.  The positions are
  $z_1=-\sqrt{2}+i\sqrt{2}, z_2=-\sqrt{2}-i\sqrt{2},
  z_3=2$.  The two figures correspond to two different choices of certain  size moduli parameters.   }}
\label{fig:fracson}
\end{center}
\end{figure}

Fractional vortex occurs also in non-Abelian gauge theories, such as the one with gauge group  $G= SO(N)\times U(1)$.    An illustrative example of fractional vortex in an $SO(6)\times U(1)$  
model is shown in Fig.~\ref{fig:fracson}.  

\section{Why non-Abelian vortices imply non-Abelian monopoles}

A more recent work of our research group concerns the monopole-vortex complex solitons occurring in systems with hierarchical gauge symmetry breaking, 
 \be
  G   \,\,\,{\stackrel {\brc \phi_{1} \ckt    \ne 0} {\longrightarrow}} \,\,\, H  \,\,\,{\stackrel {\brc \phi_{2} \ckt    \ne 0} {\longrightarrow}} \,\,\, {\mathbbm 1} \;,      \qquad   |\brc \phi_{1} \ckt|    \gg   |\brc \phi_{2} \ckt |\;.    \label{hierarchy} 
\ee
The homotopy-group sequence 
\be    \cdots \to   \pi_{2}(G)   \to    \pi_{2}(G/H)   \to      \pi_{1}(H)     \to   \pi_{1} (G) \to \cdots \,.   \label{homotopy}
\ee 
tells us that the properties of the regular monopoles arising from the breaking $G \to H$ are related to the vortices of the low-energy system. 
In particular, the fact that $\pi_{2}(G) = {\mathbbm 1}$ for any group $G$,  implies that 
\beq   \pi_{2}(G/H)   \sim      \pi_{1}(H) /  \pi_{1}(G)   \;.
\eeq
For instance, in the case of the symmetry breaking, $SU(N+1)\to SU(N)\times U(1)/{\mathbbm Z}_{N}$  the first set of checks (on the Abelian and non-Abelian magnetic flux matching)  have been done  \cite{ABEK} soon after the discovery of the non-Abelian vortex in the $U(N)$ theory.   
We wish to study more carefully the monopole-vortex configurations, taking into account a small non-BPS correction term.

For instance one might study the model based on 
hierarchically broken gauge symmetry, $SU(3) \to SU(2)\times U(1) \to  {\mathbbm 1}$,  with the Hamiltonian, 
\bea H &=&  \int d^{3}x  \Big[\, 
 \frac{1}{4 g^{2}}   (F_{ij}^{3})^{2}   +   \frac{1}{4 g_{0}^{2}}   (F_{ij}^{0})^{2}    +  \frac{1}{ g^{2}}  |{\cal D}_{i} \phi^{a}   |^{2} + 
  \frac{1}{ g_{0}^{2}}  |{\cal D}_{i}  \phi^{0} |^{2} + | {\cal D}_{i} q|^{2}    \non \\
  &   + &   g_{0}^{2} | \mu \phi^{8} + {\sqrt{2}} Q^{\dagger} t^{8} Q   |^{2} +   g^{2} | \mu \phi^{3} +  { \sqrt{2}} Q^{\dagger} t^{3} Q   |^{2}
  +  2\, Q^{\dagger}  \lambda^{\dagger} \lambda  Q  \, \Big]  \;.  \non \\
  \eea
  describing the system after the first breaking.
 Such a low-energy theory is of the  type studied in our original work on non-Abelian vortex \cite{ABEKY}, except for small terms involving $\phi^{3}(x)$ and $\phi^{8}(x)$ (which were set to their constant VEV  in that paper).  In fact, the model is the same as the one studied by Auzzi et. al. recently \cite{AEV}.  The system has unbroken, exact color-flavor diagonal $SU(2)_{C+F}$  symmetry.   Neglecting the fields which get mass of the order of the higher symmetry breaking scale, and the fields which go to zero 
exponentially outside the monopole size, 
one makes an Ansatz  (in the monopole and vortex singular gauge): 
\[     A_{\phi} =  t_{3}  A_{\phi}^{3} (\rho, z) +  t_{8}  A_{\phi}^{8} (\rho, z) ; \qquad  A_{\phi}^{3} =   -   \frac{1}{\rho}  f_{3}(\rho, z) , \quad A_{\phi}^{8} =    -  \sqrt{3}   \frac{1}{\rho}  f_{8}(\rho, z),
\]
\[ \phi ({\bf r})  =   \left(\begin{array}{ccc}  v & 0 & 0 \\0 & v & 0 \\0 & 0 & -2 v \end{array}\right) + \lambda(\rho, z) , \qquad 
 \lambda (\rho, z) =  t_{3} \lambda_{3}(\rho, z) +  t_{8} \lambda_{8}(\rho, z)\;. \]
\[ q(x) =\left(
\begin{array}{cc}
  w_1(\rho, z) & 0  \\
  0 &  w_2(\rho, z) \\
  \end{array}\right)\;,
\]
with appropriate boundary conditions.
The equations for the profile functions $f_{3}, f_{8}, w_{1} w_{2}, \lambda_{3}, \lambda_{8}$  may be studied numerically. 
Some qualitative features can be read off from the structure of these equations. \begin{description}
  \item[(i)]   The Dirac string of the monopole is hidden deep in the vortex core; the zero of the squark field at the vortex core makes 
  the singularity harmless.
  \item[(ii)]  The whole monopole-vortex complex breaks $SU(2)_{C+F}$:  the orientational zeromodes develops which live on  $SU(2)/U(1)\sim CP^{1}$.
  \item[(iii)] The  degeneracy between the monopole of the broken ``$U$ spin'' and the monopole of the broken ``$V$-spin'', which are na\"{i}vely related by the  unbroken $SU(2)$ of the high-mass-scale  breaking $SU(3) \to SU(2) \times U(1)$,   is {\it explicitly} broken  in the vacuum with small squark VEV.
    \item[(iv)]  Nevertheless, there is a new, exact continuous  degeneracy among the monopole-vortex complex configurations, related by the color-flavor symmetry  ($CP^{1}$ moduli). 
    It is possible that such an exact non-Abelian symmetry  possessed by the  monopole is at the origin of the non-Abelian dual gauge symmetry which emerges at low energies of the softly broken ${\cal N}=2$ supersymmetric QCD  \cite{CKM}. 
\end{description}

\section*{Acknowledgments} 

The main new results reported here are the fruit of a collaboration with M. Eto, T. Fujimori, S. B. Gudnason, T. Nagashima, M. Nitta, K. Ohashi,  and W. Vinci. The last part on unpublished work on the monopole-vortex complex  is based on a collaboration with S. B. Gudnason, 
D. Dorigoni, A. Michelini and M. Cipriani.  I thank them all. I wish to thank also the organizers of  the Nagoya  2009 International Workshop on ``Strong Coupling Gauge Theories in LHC Era'' [SCGT 09]  where this talk was  presented, for a stimulating atmosphere.

\end{document}